\documentstyle[aps,prl,psfig]{revtex}
\begin{document}
\draft
\title{Variational theory for site resolved protein folding free energy
	surfaces\\
      }
\author{J. J. Portman, S. Takada
\footnote{\bf present address, Department of Chemistry, 
Kobe University, Rokkodai, Kobe, 657 Japan}
and P. G. Wolynes} 
\address{Departments of Physics and Chemistry, 
University of Illinois, Urbana, IL, 61801\\  }
\maketitle

\begin{abstract}
{We present a microscopic variational theory for the 
free energy surface
of a fast folding protein that 
allows folding kinetics to be resolved to the residue level
using Debye-Waller factors as local order parameters.
We apply the method to $\lambda$-repressor and compare
with site directed mutagenesis experiments.
The formation of native structure and the free energy profile
along the folding route are shown to be well described 
by the capillarity approximation but with some fine
structure due to local folding topology.
}
\end{abstract}
\pacs{PACS number: 87.15 -v}

Proteins fold on a configurational energy landscape that has the 
shape of a funnel \cite{LMO92}.  As the protein moves down the funnel 
towards the native state, incomplete cancellation of the entropy 
and energy losses may result in free energy barriers.
So far, proteins that fold fast exhibit single
exponential kinetics \cite{Eaton97a}, consistent with
a free energy profile that has a single highest barrier along the 
progress coordinate.
Central issues are the origin of the free energy barrier for fast
folding proteins and how the ensemble of structures which represent the
bottleneck is to be characterized.
We address these questions using a variational approximation
that describes ensembles of partially folded proteins
at the highest level of resolution,
{\em i.e.,} the specific role of individual residues
in guiding the protein to the native state is quantified.
In the laboratory,
Fersht has developed a probe of the transition
state or bottleneck ensemble through protein engineering 
kinetic studies
in which the sequence of
the protein is altered by replacing residues one at a time\cite{Fersht92}. 
The experiment yields  the fraction
of the time that the mutated site is in the native conformation 
in the bottleneck ensemble by comparing folding
rates of the mutant to the wild type. Since this can be done for any 
residue in the sequence, these studies are inherently ``site resolved''.
Resolving the transition state ensemble to this level is one way to 
monitor the
average of the many routes taken as proteins fold. 

Previous analytic mean field theories and simulations have produced
energy landscapes in one or two global dimensions 
characterizing the folding ensemble\cite{PWW97}.
We develop here a free energy profile for proteins with a funneled 
landscape that is completely ``site resolved'', 
{\em i.e.,} one dimension per residue,
by extending the mean field variational calculations presented in 
\cite{SW92}.
The underlying Hamiltonian explicitly incorporates chain 
stiffness and connectivity while the approximation employs a 
variational density that monitors local order parameters for folding 
akin to the Debye-Waller factors (also called temperature factors)
for individual residues seen in X-ray crystallography.

The basic Hamiltonian for an interacting polymer chain is 
$H = H_{\rm chain} + H_{\rm int}$
where $H_{\rm chain}$ is backbone potential and $H_{\rm int}$ are
the interactions between distant monomers along the chain.
$H_{\rm chain}$ is an effective harmonic potential
$\beta H_{\rm chain} = 
1/2 \sum {\bf r}_{i} \cdot \Gamma_{ij} \cdot {\bf r}_{j}
+ B \sum {\bf r}_i^2$
where $\{{\bf r}_i\}$ are the positions of the $N$ $\alpha$-carbons, and
$\beta = 1/k_B T$ is the inverse temperature.
The first term enforces the chain connectivity while the second term
confines the  radius of gyration to a reasonable value
(achieved by fixing $B$ to a small constant: 
$B = 3/2 a^2 \times 10^{-3}$).  
For the connectivity
matrix, $\Gamma_{ij}$, we use the well known Gaussian
approximation to the
freely rotating chain derived in \cite{zwanzig78}.  
Denoting the $i^{th}$ bond vector by 
${\bf a}_i = ({\bf r}_{i+1} - {\bf r}_i)$ and the angle
between successive bond vector by $\theta$,
this stiff chain model is defined by the correlations 
$\langle {\bf a}_i \cdot {\bf a}_{i + l} \rangle = a^2 g^l$, where
$a$ is the mean bond length and $g = \cos \theta$.
Following Bixon and Zwanzig, 
$\Gamma_{ij}$ is determined by inverting these 
correlations and transforming
to the bead representation
resulting in a pentadiagonal matrix that depends 
on the stiffness parameter
$g$; (for the explicit matrix, see \cite{zwanzig78}).
In the limit $g\rightarrow 0$, $\Gamma_{ij}$ describes the standard 
flexible chain, whereas  $g\rightarrow 1$ corresponds to a rigid rod.
The persistence length, $l$, for this chain is given by $l = a/(1-g)$.
We use $g = 0.8$ giving (with $a = 3.8 {\rm \AA}$)
$l \approx 20 {\rm \AA}$, 
the persistence length for poly-L-alanine \cite{CS80}.

We take interaction between distant monomers along the chain
to be restricted to specifically native-like interactions
$H_{\rm int} = \sum \!^{(N)} \: \epsilon_{ij} u(|{\bf r}_i-{\bf r}_j|)$
where the isotropic pair potential,
has a minimum at a non-zero distance produced by summing three Gaussians,
$u(r) = \gamma_s e^{-\beta_s r^2} +\gamma_i e^{-\beta_i r^2}
 - \gamma_l e^{-\beta_l r^2}$;
the short- and intermediate-range terms are repulsive
while the long-range Gaussian is attractive
$(\beta_s > \beta_i > \beta_l)$.
The sum over pair interactions $(\sum^{(N)})$  is restricted to
native contacts.  This constraint gives a
smooth funnel shaped energy landscape, appropriate for 
fast folding proteins. This is an extreme realization of the 
{\it principle of minimum frustration} \cite{BW87} 
and is reminiscent of the lattice model originally 
introduced by G$\bar{{\rm o}}$ \cite{Go83}.
The heterogeneity of the 
interaction between different residues is reflected by the strength
$\epsilon_{ij}$.
Non-native interactions can also be included in $H_{\rm int}$
and treated by our variational method.

To study the many dimensional free energy
surface defined by $H$, we choose
local order parameters that can
characterize the ensemble of partially folded structures by 
specifying 
the temperature factor for each residue, $\hat{ {\cal B}}_i$.
This describes the mean square fluctuations of a residue 
about its native position and for fully folded proteins has
been measured.
A similar local order parameter for folding has been used
in lattice simulations \cite{pande} and earlier analytical work \cite{SW92}.
Consider the free 
energy surface defined by the set of scalar fluctuations of each residue 
from its native position  $\{ {\bf r}_i^N \}$,
$\hat{ {\cal B}}_i = ({\bf r}_i - {\bf r}_i^N)^2$.
The free energy surface ${\cal F}[\{ \hat{ {\cal B}}_i \}]$
for an ensemble specified by 
$\{\hat{ {\cal B}}_i \}$ is given by
\begin{eqnarray}
e^{-\beta {\cal F}[\{ \hat{ {\cal B}}_i \}] }
& = & {\rm Tr} \; \left[ \prod_i \delta
\left( \hat{ {\cal B}}_i - ({\bf r}_i - {\bf r}_i^N)^2 \right) 
e^{-\beta H} \right]
\\
& = & \int \!\! {\cal D} \lambda \:
{\rm Tr} \; e^{-\beta H[ \lambda]} \label{eq:f},
\end{eqnarray}
where 
$\beta H[\lambda] = \beta H 
+ \sum \lambda_i  
( \hat{ {\cal B}}_i  - ({\bf r}_i - {\bf r}_i^N)^2 )$, and
${\cal D} \lambda \equiv \prod_j d\lambda_j /2 \pi i$.

Denoting the integrand in Eq.(\ref{eq:f}) by $e^{- \beta F[\lambda]}$,
we approximate $F[\lambda]$
with the help of a reference 
Hamiltonian $H_0$ and the Gibbs-Bogoliubov variational expression
$F[\lambda]   
\approx  -k_B T \log Z_0 + \langle H[\lambda] - H_0 \rangle_0$ ,
where $Z_0 = {\rm Tr}\left[e^{-\beta H_0}\right]$, and
$\langle \cdots \rangle_0$ means the average with respect to $H_0$.
The reference Hamiltonian describes a 
Gaussian chain constrained
to fluctuate about the native structure $\{{\bf r}_i^N\}$ 
by a harmonic external field:
$\beta H_0
     = H_{\rm chain} + \sum C_i ({\bf r}_i - {\bf r}_i^N)^2$.
The variational parameters $\{C_i\}$ are conjugate
to $\{\hat{ {\cal B}}_i \}$.
This reference Hamiltonian captures the two stable phases of fast folding
proteins: the globule with small $\{C_i\}$
and the native state with uniformly large $\{C_i\}$.
$\{C_i\}$ also form a set of local order parameters
for folding.
A similar (but more elaborate) reference Hamiltonian was used
to determine the phase diagram for
proteins with a rugged energy landscape
 as well as to study folding free energy barriers\cite{SW92}.  
These mean field studies employed a global
order parameter for nativeness by setting all ${C_i}$ equal
in one region of the protein.
Different
related effective harmonic variational Hamilitonians 
have been employed to study 
polymers in random media\cite{EM88},
random directed polymers\cite{MP91},
and random copolymers \cite{MKD97}.

Evaluating Eq.(\ref{eq:f})
using the steepest descents approximation gives the stationary condition
$\hat{ {\cal B}}_i = \langle ({\bf r}_i - {\bf r}_i^N)^2 \rangle_0$ 
as a function of $\{ C_i \}$ leading to 
a one to one relation between the
$\{ \hat{ {\cal B}}_i \}$ and $\{ C_i\}$. 
Technically, it is more convenient to study the free energy 
surface in $\{ C_i \}$ space so that we consider
the variational free energy surface
expressed as $F[\{C_i\}] = E - ST$ 
with the 
estimate for the energy 
$E = \sum^{(N)} \epsilon_{ij} \langle u({\bf r}_{ij})\rangle_0$
and the entropy 
$S/k_B = \log Z_0 
+  \sum \langle C_i({\bf r}_{i}-{\bf r}^N_i)^2 \rangle_0$ 
as a function of $\{ C_i \}$.

Since $H_0$ is quadratic, $Z_0$ and all of the averages
are expressible in terms of the correlations,
$\langle {\bf r}_i \cdot {\bf r}_j \rangle_0 - 
\langle {\bf r}_i \rangle_0 \cdot \langle {\bf r}_j \rangle_0 
= 3/2 G_{ij}$, 
with 
$G_{ij} = \left[1/2\;\Gamma_{ij} + (B + C_i)\delta_{ij}\right]^{-1}$.
$Z_0$ involves the determinant of the correlation matrix while
the averages can be calculated using the density of site $i$, 
$\rho_i({\bf r}) = \langle \delta({\bf r} - {\bf r}_i) \rangle_0$,
and the pair density between sites $i$ and $j$,
$\rho_{ij}({\bf r}) = \langle \delta({\bf r} - {\bf r}_{ij}) \rangle_0$.
In terms of the average position 
${\bf s}_i = \sum_j G_{ij} C_j {\bf r}_j^N$ 
these densities are
$\rho_i({\bf r}) = 
\left( \pi G_{ii}\right)^{-3/2} 
\exp\left[-({\bf r} - {\bf s}_i)^2/G_{ii}\right]$,
and 
$\rho_{ij}({\bf r}) = 
\left(\pi \delta G_{ij}\right)^{-3/2} 
\exp\left[-({\bf r}-{\bf s}_{ij})^2/\delta G_{ij}\right]$
where $\delta G_{ij} = G_{ii} + G_{jj} - 2 G_{ij}$.

Since both $F[\{C_i\}]$ and  $\nabla_C F[\{C_i\}]$ can be expressed 
analytically in terms of $G_{ij}$ (which is calculated numerically),
it is relatively easy to locate the minima 
and saddle points numerically \cite{Wales94}.
Once the transition states and the folded, unfolded, and local minima 
are determined, we define the average folding route to be the connected
steepest descents
path from each transition state to the neighboring minima.
Only the global minimum of $F[\{C_i\}]$ is rigorously an
upper bound but the saddle points and
local minima should also be good estimates for the true free energy
surface.

We now apply the model to the folding of 
the $\lambda$-repressor protein.  $\lambda_{6-85}$
is a good candidate system since it is small 
(80 residues) and folds 
extremely rapidly in 20 $\mu$-sec following two-state kinetics \cite{Oas96}.
Recently Oas {\em et al.} probed
the structure of the transition state ensemble 
of $\lambda_{6-85}$ by comparing the folding rates measured with NMR for
seven mutants made by alanine to glycine replacements
\cite{Oas97a}.  The folding rates can be connected to the structure of
the transition state by the $\phi$ parameter developed by 
Fersht \cite{Fersht92}, 
$\phi = \Delta \log k_f /\Delta \log K$
($k_f$ is the folding rate and $K$ denotes the equilibrium constant).
$\phi \sim 0$ indicates that the conformation of the mutated residue 
in the transition state is similar to the globule, whereas 
$\phi \sim 1$ suggests that this residue has native structure in the 
transition state ensemble. Based in part on these $\phi$-values 
Table \ref{tb:phi},
Oas proposed that helices H1 and H4 are structured in the 
transition state ensemble.
While a more extensive mutation study is necessary to
characterize fully the transition state ensemble, a comparison to
these results is a strong test for the theory presented here.

We define native contacts between residues
with $\beta$-carbons within a distance of 6.5 ${\rm \AA}$
($\alpha$-carbons for glycines) in the native structure \cite{1lmbpdb} 
that are separated by at least four monomers in sequence. 
The pair distribution function of the distance between $\alpha$-carbons 
is used to constrain the 
parameters of the effective pair potential.
We find the intermediate- and long-ranged interaction parameters
$(\gamma_i,\beta_i a^2,\gamma_l, \beta_l a^2) = (9.0,0.8,6.0,0.4)$
give an effective potential well that contains all the native 
contact distances and has a minimum at the most 
probable ${\rm C}_{\alpha}-{\rm C}_{\alpha}$
contact distance,
$r^* = 1.6 a$, with $u(r^*) = -1$.
The short-range interaction represents the 
hard-core repulsion between residues and gives excluded volume;
with the choice of values
$(\gamma_s,\beta_s a^2) = (25.0,4.5)$, 
the repulsion roughly balances the attractive 
energy in the globule state allowing us to study
a folding transition that occurs directly from a random coil 
({\em i.e.,} near the theta temperature).
We will compare the free energy surface using 
a homogeneous contact strength
$(\epsilon_{ij} = \epsilon_0)$ 
with that of the full 20 letter Miyazawa-Jernigan
contact energies \cite{MJ96} in which contact between different residues
have different energies.

\begin{figure}[htb]
\hspace{.7cm}
\psfig{file=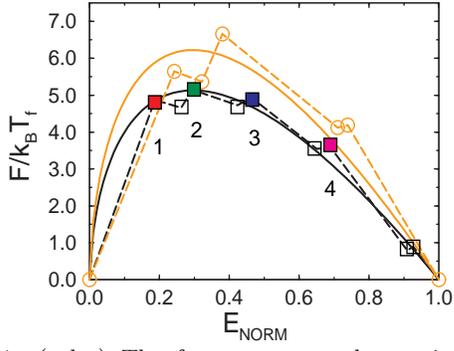,height=4.6cm,width=6cm}
\caption
{(color) The free energy at the stationary points 
along the folding route
as a function of the normalized energy for
the homogeneous (orange,\mbox{{\Large $\circ$}}) and inhomogeneous 
(black,$\Box$)
models. 
For an ensemble with average energy $E$,
$E_{NORM} = (E - E_G)/(E_N - E_G)$,
where $E_N$ and $E_G$ are the energies of the native state and 
globule state, respectively.
$f_{{\rm cap}}$ (described in the text) is also shown as the 
solid line with $\gamma = $ 35(42) $k_B T_f$ for the 
heterogeneous(homogeneous) model.}
\label{fg:fve}
\end{figure}

We now consider a low energy folding route on the free energy 
surface connecting the globule and native minima.
Fig. \ref{fg:fve} shows the 
free energy along this path 
at the folding transition temperature, $T_f$, 
plotted as a function of the fraction of energy stabilization relative
to the native state, $E_{NORM}$.
The stationary points of the free energy surface
form a broad barrier with a reasonable height $(5-7 k_B T_f)$
for fast folding proteins.
The barrier for the homogeneous case of all equal interactions 
is approximately $30\%$ 
larger than for the heterogeneous contact energy model.
Another difference between the two models is that we find
many more transition states and local minima (not shown)
for the heterogeneous case.  These arise from the
competition between contacts of different strengths.

\begin{figure}[htb]
\hspace{.7cm}
\psfig{file=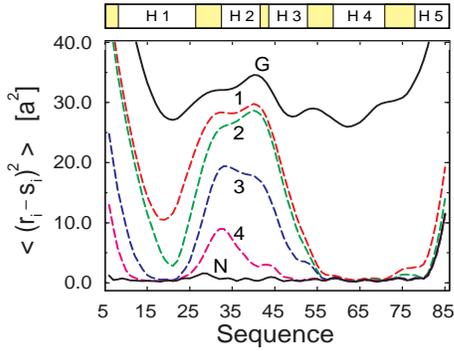,height=4.6cm,width=6cm}
\caption
{(color)  The temperature factors ({\em i.e.}, mean square fluctuations
relative to the average position of $i^{th}$ monomer, ${\bf s}_i$)
plotted as a function of sequence number
for the heterogeneous model at the stationary points 
shown in Fig.\ref{fg:fve}. 
The bar at the top indicates the helical secondary structure (H1-H5).}
\label{fg:tfac}
\end{figure}

The Debye-Waller factor (temperature factor) of each residue contains
structural information of the stationary points along the 
folding route.
The temperature factors plotted versus sequence number
at four of the saddle points along the folding route
for the heterogeneous model are shown in Fig. \ref{fg:tfac}.
Even the globule already has some structure though its fluctuations
are large.  
Comparing these curves progressively from the globule to 
the native state, we see that the barrier at $TS_1$ is described by 
the formation of helices H4-5 while the central region of
helix H1 which docks with H4 is partially localized 
but with substantial fluctuations.
This suggests that the stabilizing contacts between H4-5 and H1
are due to the general increase in density rather than any 
very strong
contacts between specific residues.
Following this is the completion of helix H5 
and the center of helix H1, 
while helices H2-3 remain relatively disordered at $TS_3$. 
Lastly, helices H2-3 become 
increasingly more ordered along this folding route as indicated by
the temperature factors at $TS_4$.

The folding route described above agrees with the conclusions
of the Oas group \cite{Oas97a}; namely, helices H1 and H4 
are structured in the transition state ensemble, 
whereas helices H2 and H3 are unstructured.
The $\phi$-values obtained
from this calculation makes the comparison more precise. 
As in the experimental analysis, we assume the ensemble of
structures do not change but recalculate the free energy for each mutant
at the saddle-points. Using $k \sim e^{-\beta \Delta F^\dagger}$,
we calculate $\phi$ at each saddle-point and their average over the
four transition states.
The results are given in Table \ref{tb:phi}.
The agreement with experiment is quite reasonable
in light of the rough approximations made in modeling the experiment.
The worst agreement is for the mutation M20. This
is a surface residue with no tertiary contacts by our definition,
thus other terms in the energy may be contributing.
Some obvious improvements to this model such as
explicit hydrogen bonding and many body forces can easily
be made, but our
aim here is to explore simplest model that give a physically
reasonable and direct picture of the folding route for 
fast folding proteins.
From this point of view, the agreement with experiment 
is very encouraging.

\begin{figure}[htb]
\hspace{.4cm}
\psfig{file=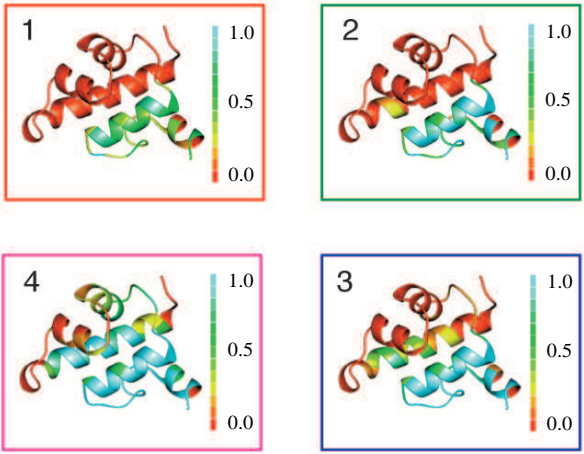,height=6.5cm,width=8.2cm}
\caption
{(color)  The 3D native structure of $\lambda$-repressor
colored according to the normalized energy
of each site, $(e_i - e_i^{G})/(e_i^{N} - e_i^{G})$
with $e_i = \sum_j^{(N)} \epsilon_{ij} \langle u( r_{ij}) \rangle_0$, 
evaluated at each saddle-point (clockwise from upper left).
$e_i^N$ and $e_i^G$ are the energies of the $i^{th}$ site in the
native state and globule state, respectively. }
\label{fg:3droute}
\end{figure}

Examination of the average folding route also leads to a simple 
physical picture for the barriers under the thermodynamic 
conditions of folding at $T_f$ directly from the random coil.
The progression of the 
folding of the 3D structure is shown in Fig.\ref{fg:3droute}, 
where the sites of the native structure
are colored according to the fraction of energy gained at that site.
The first bottleneck involves partial structure formation
in approximately 40\% of the chain (in helices H4-5). 
Subsequently, a picture much like that of the growth
of an ordered phase in an ordinary first
order transition emerges with a front of progressive ordering
crossing the protein. This is reminiscent of the capillarity theory 
\cite{BW90}.
Within the capillarity picture, one imagines an ordered region
that is completely folded separated by a sharp interface from 
a completely unfolded region.
At $T_f$, the free energy of 
progressively forming folded structure is given by
$f_{{\rm cap}} = \gamma (-N_f + N_f^{2/3})$
where $N_f$ is the fraction of native residues, and $\gamma$ is 
the surface energy cost. 
As shown in Fig. \ref{fg:fve}, 
this equation 
provides a good fit to 
the stationary points in both the homogeneous and heterogeneous models, 
identifying $N_f$ with normalized energy $E_{\rm Norm}$ (defined in 
Fig. \ref{fg:fve}) and treating $\gamma$
as a fitting parameter.

Superimposed on the average behavior of the profile are
fluctuations representing the fine structure arising from 
inhomogeneity of the local folding free energy. 
It is obvious that these fluctuations arise for the heterogeneous
model because of varying interaction energies, but
are still present for the pure homogeneous G$\bar{{\rm o}}$ like model.
This shows the high free energy intermediates \cite{pande,Olive}
along the average folding route for a very funnel-like surface
are mostly determined by the folded topology.  
Within the capillarity picture,
the smaller barrier for the heterogeneous case can be interpreted
as being due to wetting, as expected for the random field Ising
model \cite{Villain85}.  
The thermodynamic conditions studied here favor
the capillarity picture with a sharp interface.  When folding
occurs from an already collapsed state the free energy difference
of the bulk unfolded and folded phases is smaller
leading to a broader interface. 
The basic formalism can be used for this other regime as well.

We would like to thank Ben Shoemaker for helpful discussions.
This work was supported by NIH Grant No. PHS R01 GM4557.
S.T. was supported by the Japan Society for Promotion of Science.


\begin{table}[htb]
\begin{tabular}{|c|ccccccc|}
Mutant
& M15 & M20 & M37 & M49 & M63 & M66 & M81 \\ 
(Helix)
& (H1) & (H1) & (H2) & (H3) & (H4) & (H4) & (H5) \\ \hline
$\phi_{{\rm Exp}}$
&0.5 & 1.0 & 0.2 & 0.3 & 0.8 & 1.2 & 0.6 \\
$\langle \phi \rangle_{{\rm Calc}}$
&0.3 & 0.3 & 0.1 & 0.2 & 1.0 & 1.0 & 0.7 \\
\end{tabular}
\caption
{$\phi-values$ for $\lambda$-repressor.}
\label{tb:phi}
\end{table}

\begin{thebibliography}{10}

\bibitem{LMO92}
P.~E. Leopold, M. Montal, and J.~N. Onuchic, 
Proc. Natl. Acad. Sci. U. S. A. {\bf
  89},  8721  (1992);
J.~D. Bryngelson, J.~N. Onuchic, N.~D. Socci, and P.~G. Wolynes, Proteins:
  Structure, Function and Genetics {\bf 21},  167  (1995).

\bibitem{Eaton97a}
W.~A. Eaton {\it et~al.}, Curr. Opin. Struct. Biol. {\bf 7},  10  (1997).

\bibitem{Fersht92}
A.~R. Fersht, A. Matouschek, and L. Serrano, J. Mol. Biol. {\bf 224},  771
  (1992).

\bibitem{PWW97}
S.~S. Plotkin, J. Wang, and P.~G. Wolynes, J.\ Chem.\ Phys. {\bf 106},  2932
  (1997);
J.~N. Onuchic, N.~D. Socci, Z.~A. Luthey-Schulten, and P.~G. Wolynes, Folding
  and Design {\bf 1},  441  (1996);
Z. Guo, C.~L. Brooks, and E.~M. Boczko, Proc. Natl. Acad. Sci. U. S. A. {\bf
  94},  10161  (1997).

\bibitem{SW92}
M. Sasai and P.~G. Wolynes, Phys.\ Rev.\ A {\bf 46},  7979 (1992);
S. Takada and P.~G. Wolynes, Phys.\ Rev.\ E {\bf 55},  4562  (1997);
S. Takada and P.~G. Wolynes, J.\ Chem.\ Phys. {\bf 107},  9585  (1997);

\bibitem{zwanzig78}
M. Bixon and R. Zwanzig, J.\ Chem.\ Phys. {\bf 68},  1896  (1978).

\bibitem{CS80}
C.~R. Cantor and P.~R. Schimmel, {\em Biophysical Chemistry} (W. H. Freeman and
  Co., New York, 1980), Vol.~3, p.\ 1013.

\bibitem{BW87}
J.~D. Bryngelson and P.~G. Wolynes, Proc. Natl. Acad. Sci. U. S. A. {\bf 84},
  7524  (1987).

\bibitem{Go83}
N. G\={o}, Annu.\ Rev.\ Biophys.\ Bioeng.\ {\bf 12},  183  (1983).

\bibitem{pande}
V. Pande and D.~S. Rokhsar, Nature Struct.\ Biol.\  , (submitted).

\bibitem{EM88}
S.~F. Edwards and M. Muthukumar, J.\ Chem.\ Phys. {\bf 89},  2435  (1988);
J.~D. Honeycutt and D. Thirumalai, J.\ Chem.\ Phys. {\bf 90},  4542  (1989).

\bibitem{MP91}
M. M\'{e}zard and G. Parisi, J.\ Phys.\ I(France) {\bf 1},  809  (1991);
T. Garel and H. Orland, Phys.\ Rev.\ B {\bf 55},  226  (1997).

\bibitem{MKD97}
A. Moskalenko, Y.~A. Kuznetsov, and K.~A. Dawson, Physica A {\bf 249},  353
  (1998).

\bibitem{Wales94}
D.~J. Wales, J.\ Chem.\ Phys. {\bf 101},  3750  (1994).

\bibitem{Oas96}
R.~E. Burton {\it et~al.}, J. Mol. Biol. {\bf 263},  311  (1996).

\bibitem{Oas97a}
R.~E. Burton {\it et~al.}, Nature Struct.\ Biol.\ {\bf 4},  305  (1997).

\bibitem{1lmbpdb}
L.~J. Beamer and C.~O. Pabo, J. Mol. Biol. {\bf 227},  177  (1992).

\bibitem{MJ96}
S. Miyazawa and R.~L. Jernigan, J. Mol. Biol. {\bf 256},  623  (1996).

\bibitem{BW90}
J.~D. Bryngleson and P.~G. Wolynes, Biopolymers {\bf 30},  177  (1990);
A.~V. Finkelstein and A.~Y. Badretdinov, Folding and Design {\bf 2},  115
  (1997);
P.~G. Wolynes, Proc. Natl. Acad. Sci. U. S. A. {\bf 94},  6170  (1997).

\bibitem{Olive}
M. Silow and M. Oliveberg, J. Mol. Biol. {\bf 269},  611  (1997).

\bibitem{Villain85}
Villain, J.\ Phys. I(France) {\bf 46},  1843  (1985).

\end{thebibliography}
\end{document}